\documentclass[a4paper]{jpconf}
\usepackage{hyperref}
\usepackage{amsmath}	
\usepackage{amssymb}
\usepackage{color}
\usepackage{caption}
\usepackage{graphicx}
\def\xiG{\xi_G}
\def\xiW{\xi_W}
\def\xiB{\xi_B}

\newcommand{\MS}{{\ensuremath{\overline{\mathrm{MS}}}}}

\def \eps {\epsilon}

\begin{document}
\title{Three-loop beta-functions and anomalous dimensions in the Standard Model\footnote{Talk at 15th International Workshop on advanced computing and analysis techniques in physic - ACAT2013, 16-21 May 2013, Beijing, China}}
\author{\underline{A.~V.~Bednyakov}$^1$, A.~F.~Pikelner$^1$, and V.~N.~Velizhanin$^2$}

\address{$^1$ Joint Institute for Nuclear Research, 141980 Dubna, Russia}
\address{$^2$ Theoretical Physics Department, Petersburg Nuclear Physics Institute, Orlova Roscha, Gatchina, 188300 St.~Petersburg, Russia}

\ead{bednya@theor.jinr.ru}
\begin{abstract}
	In this talk	
	the methods and computer tools which were used in our recent calculation 
of the three-loop Standard Model renormalization group coefficients are discussed. 
A brief review of the techniques based on special features of dimensional regularization and
minimal subtraction schemes is given.
Our treatment of $\gamma_5$ is presented in some details. 
In addition, for a reasonable set of initial parameters the numerical estimates of the obtained
three-loop contributions are presented.

\end{abstract}

The renormalization group (RG) proves to be a useful and powerful tool in studying high-energy behavior
of the Standard Model. 
Before the discovery of the Higgs boson RG equations (RGE) were used, among other things, to bound the value of the Higgs self-coupling. 
However, the bounds significantly depends on the scale at which one expects the appearance of New Physics. 
The observation of the Higgs boson in 2012 \cite{Aad:2012tfa,Chatrchyan:2012ufa} in some sense finalizes the SM since the information about the values of all the SM couplings become available from experiments.

Due to this fact, the interest to RG studies of the Standard Model arises again, but at a new level of precision. 
One-
and two-loop results for SM beta-functions have been known for quite a long time~
\cite{SMbeta}
and are summarized in~\cite{Luo:2002ey}.

The first paper with full three-loop calculation of gauge coupling beta-functions within the SM was published in Ref.~\cite{Mihaila:2012fm}. 
The next step was carried out by another group from Karlsruhe \cite{Chetyrkin:2012rz}, which considered, in the specific limit, the three-loop beta-functions for the top quark Yukawa coupling, the Higgs self-coupling and mass parameter.
At this state of things our group entered the game. 
We not only confirmed the results of Refs.~\cite{Mihaila:2012fm,Chetyrkin:2012rz,Mihaila:2012pz} but also provided the three-loop expressions for beta-functions of all the Yukawa couplings \cite{Bednyakov:2012en} corresponding to the fermions of third generation.
Contrary to Chetyrkin and Zoller, we include the dependence on electroweak couplings.

In the beginning of 2013 our group started the calculation of missing three-loop terms in beta-functions of
the Higgs potential parameters. 
It turns out that the same problem was considered by the authors from Karlsruhe.
Unfortunately to us, they managed to obtain and make their results \cite{Chetyrkin:2013wya} public a week earlier than our group \cite{Bednyakov:2013eba}.
Nevertheless, in such a complicated calculation it is important to have a confirmation from an independent
source. 
In what follows, we are going to discuss the peculiarities of the procedure used to obtain the results published by our group in a series of papers \cite{Bednyakov:2012en,Bednyakov:2013eba,Bednyakov:2012rb}.

First of all, let us mention that the calculation of
the three-loop SM beta-functions requires evaluation of millions of Feynman diagrams. 
This task definitely requires automatization by means of a computer. 
Fortunately, all the necessary tools were available on the market, so we only needed to combine them in a proper way.

It is due to nice features of dimensional regularization and minimal subtraction scheme so one can significantly simplify the calculation. 
Since in \MS-scheme all the renormalization constants can be extracted from the ultraviolet (UV) divergent parts of corresponding Green functions, one can modify infra-red (IR) structure of the model to simplify the calculation of counter-terms. 
This is the essence of the so-called infrared-rearrangement (IRR) trick,  which was originally proposed by Vladimirov A.A~\cite{Vladimirov:1979zm}. 
This kind of modifications can lead to a spurious IR divergences which should be removed consistently by the so-called $R^*$~\cite{Chetyrkin:1984xa}. 
However, in many practical cases one can avoid this kind of complications.  In our series of paper we made use of two variants of (naive) IRR procedure. 

For the calculation of gauge and Yukawa coupling beta-functions it is possible to convert all the required two- and three- point Green functions to the massless propagator-type Feynman integrals.
It is done via neglecting all internal masses and setting the Higgs boson external momenta entering Yukawa vertex to zero. 
The evaluation of massless three-loop propagators is performed via a FORM \cite{Vermaseren:2000nd} package \texttt{MINCER}~\cite{Gorishnii:1989gt}.
This kind of manipulations does not introduce spurious IR divergences. 
Moreover, since we are only interesting in UV counter-terms it is possible to work from the very beginning within the ``unbroken phase`` of the SM, in which all the fields are massless and the Higgs doublet $\Phi$ does not have a vacuum expectation value.  

The second approach to IRR, which was used in calculation of Higgs potential parameter beta-functions, is the introduction of an auxiliary mass parameter $M$ in every propagator via iterative application of the following formula \cite{Misiak:1994zw,Chetyrkin:1997fm}
\begin{eqnarray}
	\frac{1}{(q-p)^2} = \frac{1}{q^2-M^2} 
	+ \frac{2 q p - p^2 - M^2 }{q^2 - M^2}
	\times \frac{1}{(q-p)^2}
	\label{eq:exact_dec_massless}
\end{eqnarray}
where $q$ and $p$ are linear combinations of internal and external momenta correspondingly. 
It is clear that if one applies this kind of decomposition a sufficient number of times the last term can be neglected in the calculation of UV divergences (after subtraction of subdivergences). It turns out that for the scalar four-point Green functions considered only the first term in Eq.~\eqref{eq:exact_dec_massless} is necessary.  
Consequently, we are left with massive vacuum integrals, which can be calculated by either public \texttt{MATAD} package \cite{Steinhauser:2000ry} or private \texttt{BABMA} code written by Velizhanin.
In such an approach no spurious IR divergences appear so
it can be used in the situations when a naive application of the first variant of IRR fails. 
However, the price to pay for this advantage is the necessity of explicit calculation of diagrams with counter-term insertions. This is due to the fact that one needs to introduce counter-terms for divergences contributing to the auxiliary masses for vector and scalar bosons. 
For further details see~\cite{Misiak:1994zw}.

It is worth mentioning that we can still exploit the symmetries of the unbroken SM. 
For example, all the components of the Higgs boson doublet should have  the same auxiliary mass counter-term. 
The same is true for the SU(2) gauge bosons.  

Moreover, as it is stated in Ref.~\cite{Chetyrkin:1997fm}, the auxiliary mass appearing in the numerator in RHS of Eq.~\eqref{eq:exact_dec_massless} can be safely neglected since it can only contribute to the unphysical mass counter-terms. 
In addition, one can also skip Feynman diagrams with vacuum subdiagrams. In spite of the fact that these subdiagrams are non-zero when the auxiliary mass is introduced, they still can be neglected due to the same reasons.

As it was noticed above there are a lot of diagrams which should be generated and evaluated in order to 
find three-loop contributions to the considered quantities. 
In our calculation we made use of two popular
codes, \texttt{FeynArts}\cite{Hahn:2000kx} and \texttt{DIANA} \cite{Tentyukov:1999is}/\texttt{QGRAF} \cite{Nogueira:1991ex}, which generate necessary diagrams and produce corresponding analytic expressions. 
Both packages require a model file prepared in a special format to do their job.
Since we were wanting to simplify the calculation as much as possible we prepared a model file for the unbroken SM in the background field gauge (BFG) \cite{Abbott:1980hw,Denner:1994xt}. 
This kind of gauge allows one to find gauge coupling beta-functions solely from UV-divergences of corresponding gauge field propagators. We used a very fast \texttt{LanHEP} code \cite{Semenov:2010qt} by A. Semenov  to derive all the SM vertices from the considered SM Lagrangian (see.~\cite{Bednyakov:2012rb}) in \texttt{FeynArts} notation.  
It is worth mentioning that the Karlsruhe group made use of alternative package \texttt{FeynRules} \cite{Christensen:2008py} to solve similar problem.
Latter on a simple script was written to convert the \texttt{FeynArts} model file to that of \texttt{DIANA}.

 A typical problem which arises in this kind of calculation is internal momenta identification.
 In order to evaluate a Feynman diagram one needs to use the momenta notation of the chosen code (\texttt{MINCER}/\texttt{MATAD}/\texttt{BAMBA}). 
 In the case of gauge and Yukawa couplings the problem was solved with the help of routine, \texttt{MapMincer}, which associate with every \texttt{FeynArts} topology the corresponding \texttt{MINCER} topology and distribute \texttt{MINCER} momenta accordingly. 
 It is worth mentioning that \texttt{MapMincer} can deal with three-point vertices with one external leg carrying zero momentum (i.e., when internal lines have dots).
 The corresponding routine for \texttt{DIANA}/\texttt{QGRAF}, \texttt{MapDiana}, performs similar task, but maps every generated topology to fully massive vacuum integrals, which appear after the mentioned``exact'' decomposition \eqref{eq:exact_dec_massless} of internal propagators\footnote{Both \texttt{MapMincer} and \texttt{MapDiana} are coded by A.~Pikelner.}.

\begin{figure}[h]
\begin{center}
	\includegraphics[width=30pc]{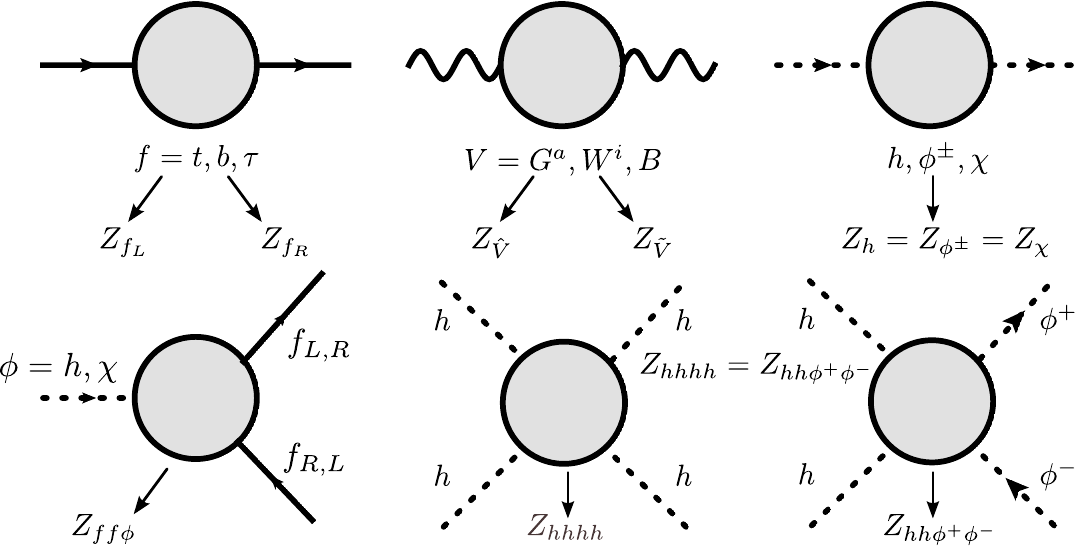}
\end{center}
\caption{\label{fig:1}The Green functions, considered in the beta-function calculations, together with the corresponding renormalization constants. Left- and right-handed fermions, denoted by $f_{L,R}$, renormalise differently in the SM. The same is true for background $\hat{V}$ and quantum $\tilde{V}$ gauge fields in BFG employed. The SU(2) invariance implies the presented equalities, which serve as an additional cross-check.}
\end{figure}

 Given a model file for \texttt{FeynArts}/\texttt{DIANA} together with the correct mapping of internal momenta to the notation of the utilized three-loop codes it is tenuous but straightforward to generate and calculate one-, two-, and three-loop contributions to the 1PI Green-functions presented in Fig.~\ref{fig:1}. 
 For the SU(3) color algebra the \texttt{COLOR}~\cite{vanRitbergen:1998pn} package was used.

From two-point Green functions we extract the corresponding wave function renormalization constants in the \MS-scheme. 
It is worth pointing that the self-energies of both background and quantum gauge fields are considered. 
The renormalization constants of the former, $Z_{\hat V}$, are directly connected to that of gauge couplings and the renormalization of the latter corresponds to the Z-factors of three gauge-fixing parameters, $\xiG$, $\xiW$, and $\xiB$, which we keep non-zero during the whole stage of calculation. The independence of the final results on these parameters serves as an important cross-check of the obtained expressions.

The three-point 1PI functions for the Yukawa vertices with neutral Higgs bosons $h$ and $\chi$ are also presented in Fig.~\ref{fig:1}. It is interesting to note that our calculation explicitly demonstrated that the semi-naive treatment of $\gamma_5$ discussed below is only applicable if one takes into account the gauge anomaly cancellation condition $N_C=3$. Here $N_C$ denotes the number of colors.

The beta-functions of the SM parameters are extracted from the corresponding renormalization constants. For the
gauge and Yukawa couplings we used the following relations 
\begin{equation}
	Z_{g_1} = Z_{\hat B_1}^{-1/2},\qquad  	
	Z_{g_2} = Z_{\hat W}^{-1/2},\qquad  	
	Z_{g_s} = Z_{\hat G}^{-1/2}, \qquad 
	Z_{y_f} = \frac{Z_{ff\phi}}{\sqrt{Z_{f_L} Z_{f_R} Z_\phi}},
	\label{eq:gauge_yukawa_rc}
\end{equation}
	where $g_1$, $g_2$  are SU(2) and U(1) gauge couplings correspondingly, $g_s$ denotes the strong coupling, and $y_f$ is the Yukawa coupling associated with the (right-handed) fermion $f=t,b,\tau$. 
Both neutral components, $\phi=h,\chi$, gave the same result, and, thus, provided us with a confirmation of the validity of the obtained expressions. 
For the Higgs self-coupling it is impossible to use \texttt{MINCER} naively, so Feynman diagrams for
	the four-point functions (see Fig.~\ref{fig:1}) converted to the fully massive vacuum integrals
	were calculated with the help of private code \texttt{BAMBA} 
	(by Velizhanin, who considered the $hh\phi^+\phi^-$ vertex) and public package 
	\texttt{MATAD} (by Bednyakov and Pikelner who considered the fully symmetric $hhhh$ vertex).  
These two independent evaluations lead to the same final expression for the vertex 
	renormalization constants, i.e., confirming the SU(2) relation $Z_{hhhh} = Z_{hh\phi^+\phi^-}$.

\begin{figure}[h]
\begin{center}
	\includegraphics[width=30pc]{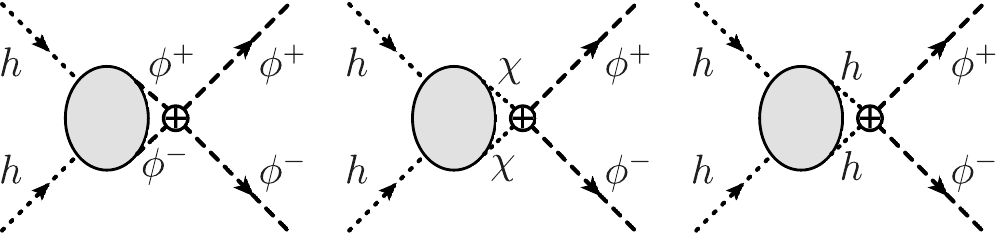}
\end{center}
\caption{\label{fig:2}The restricted set of Feynman diagrams which is used to obtain the renormalization constant $Z_{hh[\phi^+\phi^-]}$. The diagrams of the first type have to be multiplied by $1/2$.}
\end{figure}

A comment is in order on the Higgs mass parameter $m^2$.
It is possible to obtain the corresponding anomalous dimension by considering
the renormalization of the $|\Phi|^2$ composite operator within the unbroken(=massless) SM (see, e.g., \cite{Chetyrkin:2012rz}). 
This kind of result can be found at almost no cost from the calculation of $hh\phi^+\phi^-$ vertex.
It is sufficient to select the diagrams, which have $\phi^+$ and $\phi^-$ external fields connected to the same four-point vertex (see Fig.~\ref{fig:2}), and weight different contributions with a correct combinatorial 
factor. This restricted set of diagrams give rise to the $Z_{hh[\phi^+\phi^-]}$  renormalization constant.

At the end of the day, the renormalization constants for $\lambda$ and $m^2$ are obtained with the help 
of the following relations
\begin{equation}
	Z_\lambda = \frac{Z_{hhhh}}{Z_h^2} = \frac{Z_{hh\phi^+\phi^-}}{Z_h Z_{\phi^\pm}}, 
	\qquad
	Z_{m^2} = \frac{Z_{hh[\phi^+\phi^-]}}{Z_{h}}
	\label{eq:lambda_mass_RC}
\end{equation}

From renormalization constants $Z_{a_i}$ for the dimensionless SM parameters,  
\begin{equation}
	 a_i   =   \left(\frac{5}{3} \frac{g_1^2}{16\pi^2}, \frac{g_2^2}{16\pi^2}, \frac{g_s^2}{16\pi^2},
	 \frac{y_t^2}{16\pi^2}, \frac{y_b^2}{16\pi^2}, \frac{y_\tau^2}{16\pi^2}, \frac{\lambda}{16\pi^2}\right),
	\label{eq:coupl_notations}
\end{equation}
	it is straightforward to obtain the corresponding beta-functions 
\begin{equation}
	\beta_i(a_k) = \frac{d a_i(\mu,\epsilon)}{d \ln \mu^2}\bigg|_{\epsilon=0},\quad 		
 	\beta_i =  \sum_{l} a_l\frac{\partial c_i^{(1)}}{\partial a_l} - c_i^{(1)},
	\quad
	a_{i,\mathrm{Bare}}\mu^{-2\epsilon} = Z_{a_i} a_i  = a_i +\sum_{n=1}^\infty \frac{c_i^{(n)}}{\epsilon^n} 
	 	\label{eq:beta_calc_simple}
\end{equation}
Here $\mu$ is the \MS~renormalization scale, $\epsilon = (4-D)/2$ is the parameter of dimensional regularization, and $c_i$ denotes the coefficient for the single pole in $\epsilon$ in the expression for $Z_{a_i}$, which
enters the relation between the bare parameters $a_{i,\mathrm{Bare}}$ and the renormalized ones.

For the anomalous dimension of the Higgs mass parameter $m^2$ one can use similar formulae
\begin{equation}
	\gamma_{m^2}(a_k) = \frac{d \ln m^2 }{d \ln \mu^2}\bigg|_{\epsilon=0} = 
	\sum_{l} a_l\frac{\partial c_{m^2}^{(1)}}{\partial a_l}, 
	\qquad
	m^2_{\mathrm{Bare}} = Z_{m^2 } m^2  = m^2 \left( 1 +\sum_{n=1}^\infty \frac{c_{m^2}^{(n)}}{\epsilon^n} \right) 
	 	\label{eq:anom_calc_simple}
\end{equation}
	The full analytical results for the considered quantities can be found in ancillary files of the arXiv versions of our papers. The intermediate expressions, e.g., all the renormalization constants, can also be obtained, if needed, from the authors. 
	It is worth mentioning that the beta-functions of all the fundamental SM parameters are free from gauge--parameter dependence, which is a crucial test for our calculation. 
	In addition, the anomalous dimension of the Higgs doublet can be of some interest, so we also include the corresponding expression in the ancillary files of Ref.~\cite{Bednyakov:2013eba}. 
	
	Before going to the results we would like to stop on the problem related to the definition of  the $\gamma_5$ matrix withing the dimensional regularization.  
It is known from the literature (see, e.g., Ref.~\cite{Jegerlehner:2000dz} and recent explicit 
	calculation~\cite{Chetyrkin:2012rz}) that the traces with an odd number of $\gamma_5$ appearing for the first time
	in the three-loop diagrams require special treatment.
We closely follow the semi-naive approach presented in Refs.~\cite{Chetyrkin:2012rz,Mihaila:2012pz}.
First of all, we anticommute $\gamma_5$ to the rightmost position in a fermion chain and use $\gamma_5^2 = 1$.
In the ``even'' traces all $\gamma_5$ are contracted with each other, so the corresponding expressions can
	be treated naively in dimensional regularization.
In ``odd'' traces we are left with only one $\gamma_5$.
These traces are evaluated as in four dimensions and produce totally antisymmetric tensors via the relation
\begin{eqnarray}
	\tr\left( \gamma^\mu \gamma^\nu \gamma^\rho \gamma^\sigma \gamma_5 \right) = -4 i \epsilon^{\mu\nu\rho\sigma}
	\label{eq:gamma5_trace}
\end{eqnarray}
	with $\epsilon^{0123} = 1$.

Due to the fact that we use both the $\gamma_5$ anticommutativity and 
	the four-dimensional relation \eqref{eq:gamma5_trace}, the cyclicity of the trace should be relinquished
	\cite{Korner:1991sx}.
One has to choose a certain ``reading prescription'' for an ``odd'' Dirac trace, i.e., start reading a closed fermion chain from a proper place, in order to achieve the correct final result. 
However, in our calculations the problem of $\gamma_5$ positioning within the ``odd'' traces is
	solved implicitly since the diagram generation routines split the traces for us at certain points.
It should be stressed that a non-trivial contribution to the considered quantities can only appear when there are two ``odd'' traces in a diagram, since two $\epsilon$-tensors should be ``contracted'' with each outher to produce a non-zero effect. 
We can distinguish two situations.
Two ``odd'' traces in our three-loop calculations can appear either as two internal fermion loops in a diagram with external bosons or, if one considers Green functions with two external fermions,  one Dirac trace from internal fermion loop can be combined with the trace appearing after contraction with an appropriate projector.   

It is easy to convince oneself that in the case of two internal traces only triangle subloops with three external vector particles can potentially produce ``eps''-tensors. 
However, it is known that in the SM these kind of traces cancel upon summation over all the fermion species due to the absence of gauge anomalies~\cite{Bouchiat:1972iq,Gross:1972pv}. The same is true if one considers fermion self-energies up to three-loops. 

\begin{figure}[h]
	\begin{center}
\includegraphics[width=14pc]{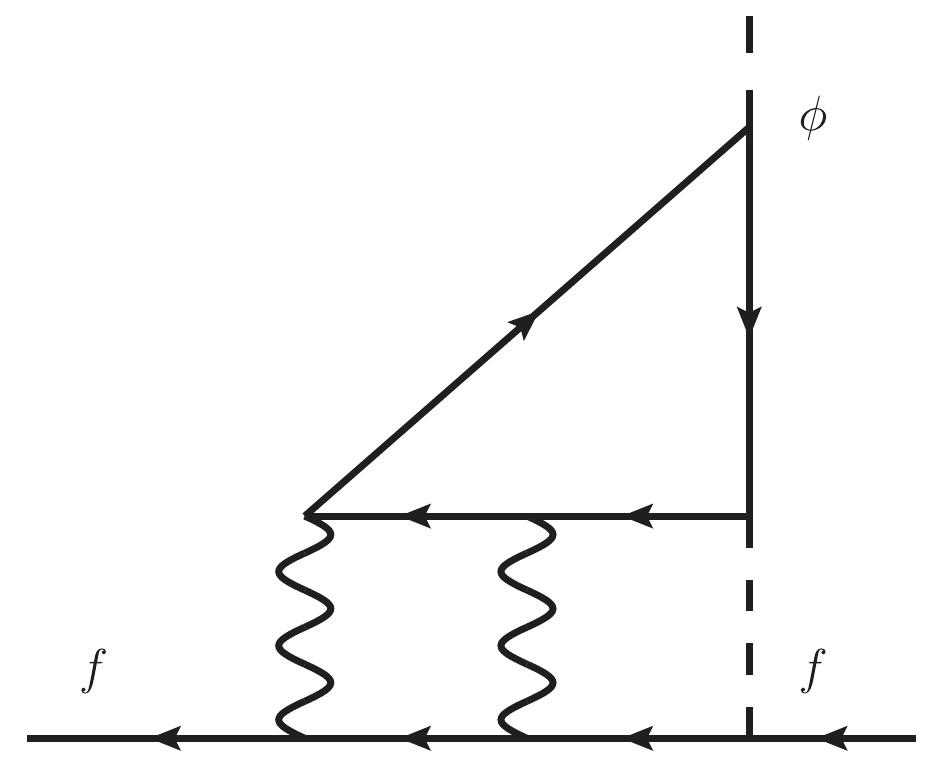}\hspace{2pc}%
\begin{minipage}[b]{18pc}\caption{\label{fig:3} The diagrams for the fermion-fermion-Higgs vertex ($ff\phi$), which produce a non-zero contribution due to contraction of two $\eps$-tensors appearing from two Dirac traces with $\gamma_5$. The ambiguity in positioning of $\gamma_5$ within the traces does not affect the UV-divergent part.}
\end{minipage}
\end{center}
\end{figure}

It turns our that the non-trivial contribution due to the contraction of two $\eps$-tensor appears from the Yukawa vertex  (see Fig.~\ref{fig:3}). This kind of diagrams was also considered in \cite{Chetyrkin:2012rz}.

A final remark about $\gamma_5$ is again related to the fact that the ambiguity in the choice of ``reading'' point in the ``odd'' traces can only spoil our result for the three-loop UV-divergence in the case of the mentioned triangle subgraphs, for which one can set all the ``odd'' traces to zero from the very beginning. 
For the only non-trivial case all the ``reading'' points are equivalent since the difference gives the contribution $\mathcal{O}(\eps^0)$ which we neglect here.

\begin{figure}[th]
\begin{center}
	\includegraphics[width=\textwidth]{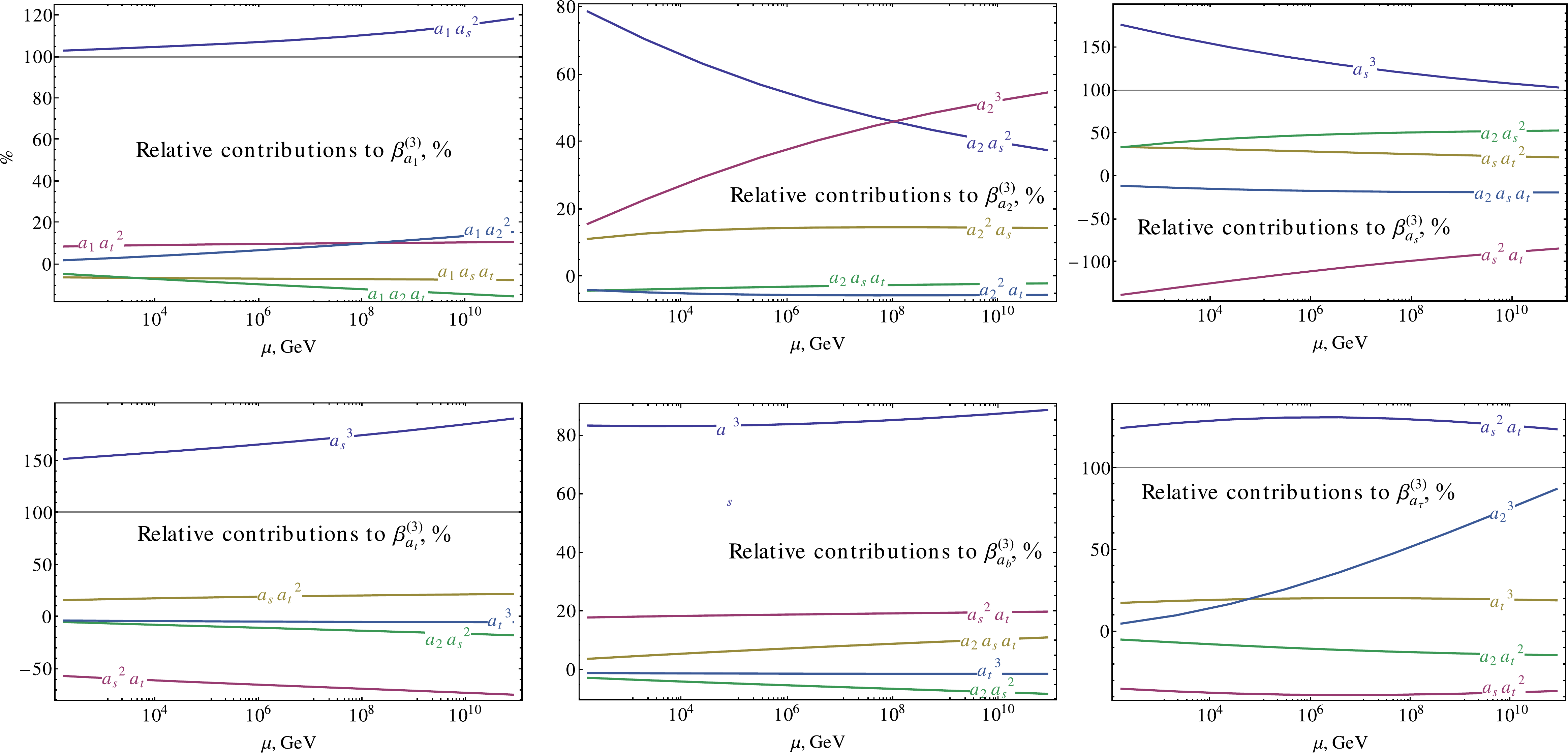} 
\end{center}
\caption{\label{fig:4} The scale dependence of the relative contributions of different terms in the three-loop corrections to the beta-functions of the (squared) gauge and Yukawa couplings. Only the most sizable corrections are shown. The boundary conditions at scale $\mu=100$~GeV are given in \eqref{eq:in_values}. }
\end{figure}

It is obvious, that the resulting expression for the three-loop contributions to the considered renormalization group quantities are too lengthy to be presented here. 
For the demonstration purposes only, we would like to show how different relative contributions to the three-loop corrections to the SM coupling beta-functions evolve with renormalization scale $\mu$ (see Fig.~\ref{fig:4}). 
For a reasonable choice of the SM initial running parameters\footnote{obtained with the help of F.~Bezrukov code~\href{http://www.inr.ac.ru/~fedor/SM}{http://www.inr.ac.ru/\~{}fedor/SM}} at the  scale $\mu=100$ GeV 
\begin{equation}
	g_1 = 0.3576,~g_2 = 0.6514,~g_s = 1.2063,~y_t = 0.9665,~y_b = 0.016,~y_\tau = 0.01,~\lambda = 0.13
	\label{eq:in_values}
\end{equation}
we solve the corresponding RGE numerically up $5\cdot 10^{10}$ GeV. 
With the help of these solutions one can evaluate the three-loop beta-functions at any scale and find how different terms contribute to to the total value of $\beta^{(3)}_i$. 
From Fig.~\ref{fig:4} one can see that the dominant contributions is due to the strong and top Yukawa couplings.
However, with the increase of the renormalization scale the SU(2) coupling can also play a role.

\begin{figure}[htb]
\begin{center}
\begin{tabular}{cc}
	\includegraphics[width=0.45\textwidth]{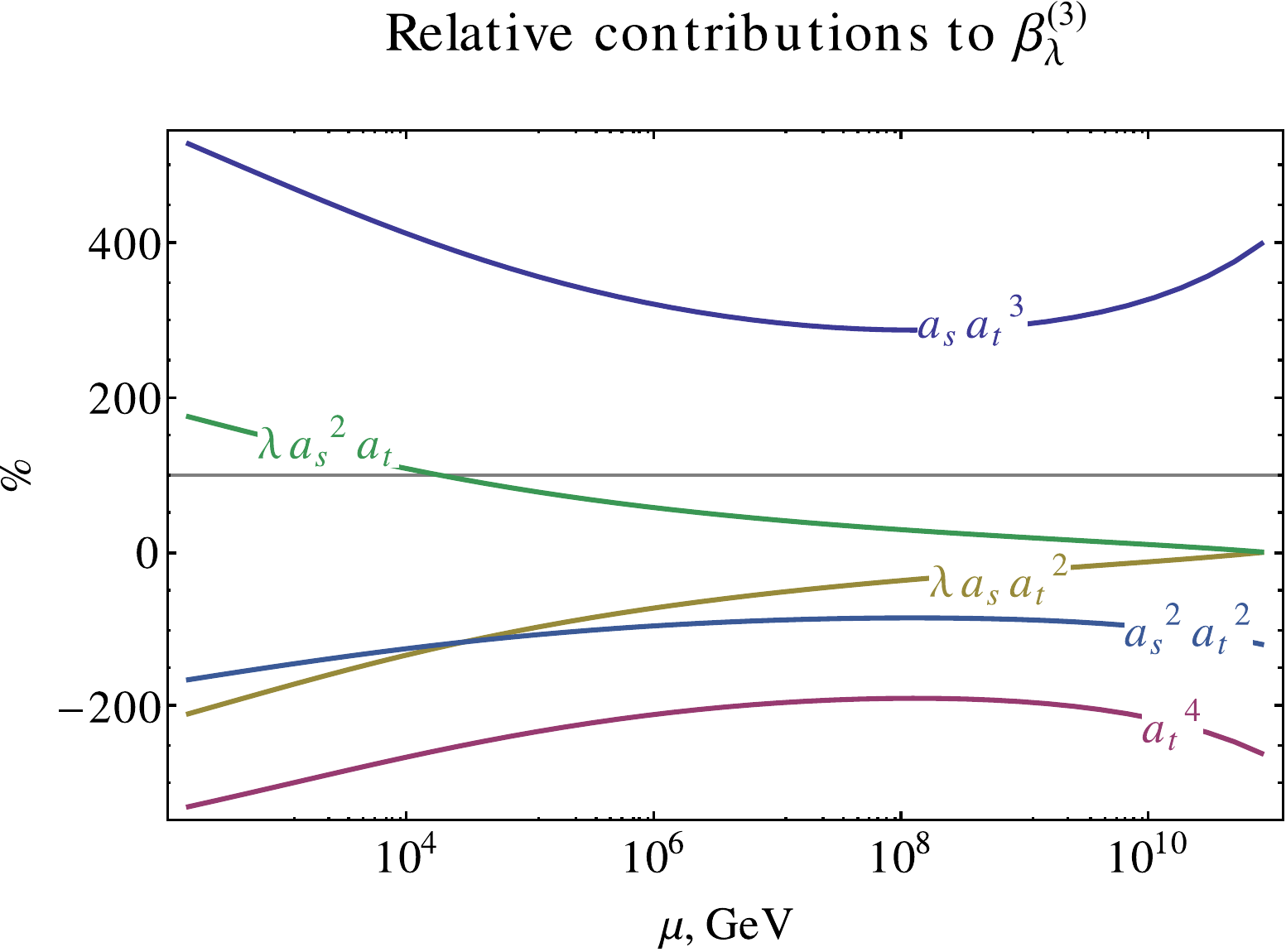} & 
	\includegraphics[width=0.45\textwidth]{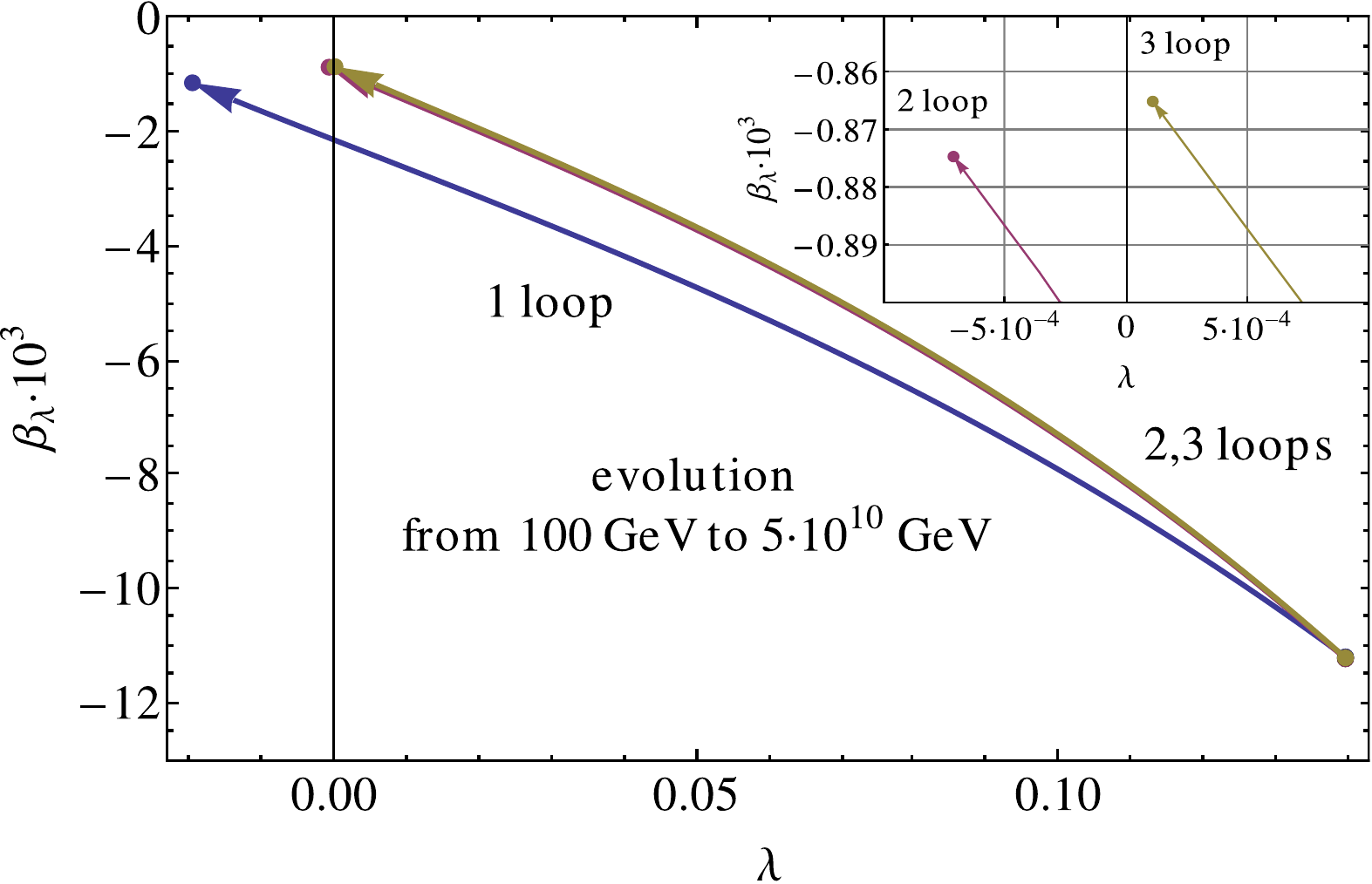}  \\
\end{tabular}
\end{center}
\caption{\label{fig:5} The scale dependence of the relative contributions to the three-loop corrections $\beta^{(3)}_{\lambda}$. Only the most sizable corrections are shown. Phase space trajectories form 1-,2-,and 3-loop evolution are provided in the plane $(\lambda, \beta_{\lambda})$ The boundary conditions are given in Eq.~\eqref{eq:in_values} }
\end{figure}

It is fair to say that the most interesting SM beta-function is that of the Higgs self-coupling, since from the evolution of the latter one can deduce the so-called ``vacuum stability bound''~(see recent papers~\cite{Bezrukov:2012sa,Degrassi:2012ry,Alekhin:2012py} and references therein).
In Fig.~\ref{fig:5} one can find the same evolution of the relative contributions to $\beta_\lambda^{(3)}$.
In addition, the slice $(\lambda, \beta_\lambda)$ of the phase space $(a_i, \beta_{a_i})$ is presented together with trajectories, obtained with the help of one-, two-, and three-loop evolution from 100 GeV to $5\cdot 10^{10}$ GeV. 
One can see that for the given set of initial conditions the running $\lambda(\mu)$ is driven to zero faster when one-loop RGEs are employed instead of two- or three-loop ones.
The difference between two- and three-loop evolution is not sizable, but the fact that with three-loop corrections the scale at which $\lambda$ hits zero slightly higher, favours the latter.

To conclude, we obtained the three-loop beta-functions for the SM parameters. The results for gauge and Higgs-potential couplings coincide with that obtained by two Karlsruhe groups. The beta-functions for Yukawa couplings were obtained for the first time.   
Moreover, we established a framework that allow us to carry out a similar calculation within an ``arbitrary'' QFT model. 
However, it should be stressed that in a self-consistent RGE analysis of the chosen model
the obtained RGEs should be accompanied by the so-called threshold (matching) corrections (see, e.g.,Refs.~\cite{Bezrukov:2012sa,Jegerlehner:2012kn,Degrassi:2012ry} for the recent SM results).
\ack
AVB is grateful to the organizers and conveniers of the ACAT2013 workshop for the invitation and 
for arranging such a nice event. In addition, we would like to thank M.~Kalmykov for drawing our attention to the problem and stimulating discussions.  
The work is supported in part by RFBR grants 11-02-01177-a, 12-02-00412-a, and by JINR Grant No. 13-302-03. 
\section*{References}
\providecommand{\newblock}{}

\end{document}